# EPSILON: Adaptive Fault Mitigation in Approximate Deep Neural Network using Statistical Signatures

Khurram Khalil, Khaza Anuarul Hoque
*Department of Electrical Engineering and Computer Science*
*University of Missouri-Columbia, USA*
{khurram.khalil, hoquek}@missouri.edu

*Abstract*—The increasing adoption of approximate computing in deep neural network accelerators (AxDNNs) promises significant energy efficiency gains. However, permanent faults in AxDNNs can severely degrade their performance compared to their accurate counterparts (AccDNNs). Traditional fault detection and mitigation approaches, while effective for AccDNNs, introduce substantial overhead and latency, making them impractical for energy-constrained real-time deployment. To address this, we introduce EPSILON, a lightweight framework that leverages pre-computed statistical signatures and layer-wise importance metrics for efficient fault detection and mitigation in AxDNNs. Our framework introduces a novel non-parametric pattern-matching algorithm that enables constant-time fault detection without interrupting normal execution while dynamically adapting to different network architectures and fault patterns. EPSILON maintains model accuracy by intelligently adjusting mitigation strategies based on a statistical analysis of weight distribution and layer criticality while preserving the energy benefits of approximate computing. Extensive evaluations across various approximate multipliers, AxDNN architectures, popular datasets (MNIST, CIFAR-10, CIFAR-100, ImageNet-1k), and fault scenarios demonstrate that EPSILON maintains 80.05% accuracy while offering 22% improvement in inference time and 28% improvement in energy efficiency, establishing EPSILON as a practical solution for deploying reliable AxDNNs in safety-critical edge applications.

## I. INTRODUCTION

The popularity of Deep Neural Networks (DNNs) in edge computing and IoT devices has led to increased adoption of approximate computing techniques to address energy efficiency challenges. Approximate DNNs (AxDNNs) leverage the inherent error tolerance of neural networks to achieve significant energy savings [1]. However, in safety-critical scenarios such as autonomous vehicles, healthcare monitoring, and industrial control systems, permanent faults in AxDNN accelerators pose severe reliability risks. Recent studies show that permanent faults in AxDNNs can cause up to 66% accuracy loss, compared to only 9% in accurate DNNs (AccDNNs) for the same faulty bit [2]. Hence, it is crucial to develop just-in-time (JIT) techniques for fault detection and mitigation in approximate computing environments that must maintain continuous system operation without interruption and introduce minimal overhead to preserve the energy benefits of approximate computing.

The existing research for detecting faults in AI accelerators (and other computing systems) primarily focuses on conventional approaches like functional testing, Built-in-self-test (BIST), and error correction codes (ECCs) [3], [4], [5]. However, these techniques face unique challenges in approximate computing environments. The functional testing strategies require periodic execution of test patterns; hence they are not JIT compliance. Similarly, BIST and ECCs introduce significant overhead, often negating the energy efficiency benefits of approximate computing [4], [6]. Manufacturing test methods like Automatic Test Pattern Generation (ATPG) require switching between functional and test modes, making them impractical for JIT fault detection in approximate systems. On top of that, these approaches are not designed to effectively distinguish between intentional approximations and actual faults. For mitigation, current approaches primarily rely on techniques like DNN retraining and hardware redundancy [2], [7]. However, retraining-based solutions require access to training data and introduce substantial computational overhead, making them impractical for edge deployment. Furthermore, traditional redundancy-based fault tolerance techniques like dual-modular redundancy (DMR)/triple-modular redundancy (TMR), while effective for accurate computing, introduce excessive area and power overhead in approximate computing scenarios [4]. These limitations highlight the need for specialized fault detection and mitigation techniques that can preserve the benefits of approximate computing while ensuring their reliable operation.

Multi-exit neural networks (MENNs) have emerged as a promising solution for addressing these challenges due to their inherent flexibility and adaptability. MENNs allow early predictions by introducing intermediate exits at various layers of a DNN, enabling faster inference and improved energy efficiency [8], [9], [10]. Any standard DNN architecture can be converted into a multi-exit variant by strategically placing auxiliary classifiers or regression heads at selected layers. This adaptability makes MENNs highly versatile for diverse applications. Many recent advancements in MENNs, such as HyperGCN [11], DE3-BERT [12], EENet [13], Adadet [14], BranchyNet [10] and SDN [8], demonstrate their ability to achieve significant reductions in computation time and energy consumption without compromising accuracy. Additionally, MENNs offer robust fault tolerance capabilities, as their distributed decision-making process inherently isolates errors to specific branches, minimizing their impact on overall performance [15], [16]. Beyond these advantages, MENNs also mitigate several common challenges in deep learning, including speeding up the inference process, alleviating the vanishing gradient problem, reducing overfitting and overthinking tendencies, and supporting DNN partitioning across devices [17], [18], [8]. These features make them ideal for multi-tier computation platforms, such as edge computing environments, where resource constraints and latency requirements are critical [19], [20], [21]. Owing to these compelling benefits, MENNs are gaining increasing popularity in the research community, with applications spanning both traditional DNN architectures and modern transformer-based models [22], [23], [24]. These advantages make MENNs particularly well-suited for fault detection and mitigation in approximate computing environments, where maintaining both energy efficiency and reliability is paramount.

In this paper, we introduce EPSILON, an efficient fault detection and mitigation framework designed explicitly for AxDNNs. Our method leverages MENN architectures [8] combined with statistical signatures to effectively distinguish between acceptable approximations and permanent faults, eliminating the need for conventional test patterns. Central to our approach is a layer-importance aware detection mechanism that dynamically adjusts detection thresholds based on both positional and structural importance of layers, ensuring robust fault detection while preserving approximate computing benefits. The framework implements a two-stage strategy that first utilizes exit confidence monitoring for efficient processing, activating sophisticated statistical pattern analysis only when necessary, thus maintaining computational efficiency while providing theoretical error bounds. A key advantage of our proposed method is its ability to operate effectively without prior knowledge of fault locations. The key contributions of this paper are summarized as follows.

- We propose EPSILON, a framework designed for JIT fault detection in AxDNNs. By utilizing multi-exit architectures, EPSILON ensures built-in fault tolerance and supports early termination. The framework features a lightweight, runtime statistical signature detection method that operates without needing training data, making it highly suitable for deployment in edge environments.
- We develop a layer-importance aware fault mitigation mechanism that dynamically adjusts affected neurons based on statistical patterns and layer criticality, achieving consistent performance across different fault rates (10% –50%) and network architectures.
- We evaluate EPSILON comprehensively across multiple AxDNN architectures, datasets, and approximate multipliers, maintaining up to **80.05% average accuracy under severe fault conditions (50% fault rate), a 22% reduction in inference time, and up to 28% improved energy efficiency compared to state-of-the-art**.

The remainder of this paper is organized as follows: Section 2 provides background on approximate computing and reviews related work in fault tolerance for AccDNNs. Section 3 details our proposed EPSILON framework and its components. Section 4 presents our experimental methodology and results. Finally, Section 5 concludes the paper and discusses future research directions.

## II. BACKGROUND

### A. Statistical Fault Detection

Conventional fault tolerance methods, such as Error Correction Codes (ECCs) and Triple Modular Redundancy (TMR), often impose significant overheads, undermining the advantages of approximate computing [25], [26]. A comprehensive list of such techniques for AccDNN fault detection and mitigation can be found in [27], [28]. There are also recently proposed AccDNN fault repair methods that suffer from the burden of retraining [29], [30], [2]. In contrast to these methods, advances in uncertainty quantification offer more efficient alternatives, including energy-based methods [31], Bayesian DNNs [32], and Monte Carlo Dropout [33], which can effectively distinguish between observed and unobserved data distributions. These techniques leverage metrics, such as the energy score, epistemic and aleatoric uncertainty, etc., that typically exhibit elevated values for anomalous inputs that may have been caused by hardware faults, enabling precise fault detection.

### B. Multi-Exit Neural Networks

Multi-exit neural networks (MENNs) represent a significant advancement in DNN architecture design, offering both efficiency and inherent fault tolerance capabilities [8]. For an approximate MENN $F$ with $N$ exits, each exit $i$ provides a prediction for input $x$, enabling early termination when sufficient confidence is achieved. Formally, the network terminates at exit $i$ if $\max(F_i(x)) > \gamma$, where $\gamma$ is a predetermined confidence threshold. This architecture is particularly relevant for fault-tolerant computing as it provides multiple redundant paths for prediction, allowing the network to maintain reliability even when certain exits are affected by hardware faults. Building on these statistical methods and MENN principles, our research introduces a novel JIT fault detection strategy that optimally balances energy efficiency and system reliability in AxDNNs.

## III. PROPOSED EPSILON FRAMEWORK FOR AxDNN FAULT DETECTION & MITIGATION

This section introduces EPSILON, an efficient framework for fault detection and mitigation in AxDNNs, emphasizing its adaptability to approximate computing scenarios. The framework is divided into *pre-deployment* and *inference phase*, described in coming sub sections.

### A. Problem Formulation

The primary objective of the EPSILON framework is to detect and mitigate faults within AxDNN while ensuring minimal energy consumption and computational overhead. Formally, let $x$ denote the input data to the AxDNN, and $F$ represent the approximate multi-exit AxDNN architecture comprising exits $F_i$, where $i$ spans the total number of exits $N$. The framework implements a predefined confidence threshold $\gamma$ to decide if deeper inspection is necessary based on the confidence levels of predictions at each exit. The statistical foundation of the framework is established through layer-specific signatures $S_l$ for layers $l = 1$ to $L$, where each signature $S_l$ incorporates four fundamental statistical measures: mean $\mu_l$, standard deviation $\sigma_l$, quartiles $Q_l$, and sparsity pattern $\rho_l$. Layer-specific importance factors $\alpha_l$ describe a set of importance factors for each layer, which influences fault detection thresholds and corrective actions. For the $i^{th}$ exit, the network generates prediction $pred_i$ with an associated confidence score $conf_i$. The framework utilizes candidate valid weight values $v$ derived from $Q_l$, while $W_l$ denotes the weights for layer $l$. The fault detection threshold $T_l$ is dynamically computed as a function of $\sigma_l$ and layer importance factor $\alpha_l$, incorporating an offset factor $m$. The fault position (FP) identifies the layer at which faults manifest, enabling precise fault localization and subsequent analysis.

### B. Pre-Deployment Phase

The pre-deployment phase establishes the foundation for fault detection through its key components. In our framework, fault-free multi-exit AxDNNs act as the golden model. This baseline provides the reference to derive statistical metrics and set confidence thresholds for effective fault detection. By analyzing these fault-free networks, we generate *statistical signatures* that characterize each layer's normal behavior. These signatures guide our detection mechanisms, ensuring that runtime deviations are accurately identified as anomalies. The confidence threshold $\gamma$ is predetermined as outlined in [8], [16]. Next, we describe the *signature generation* process, followed by our *layer-importance-aware detection* mechanism and the *adaptive mitigation* strategy.

*1) Statistical Signature Generation:* The statistical signatures in EPSILON are computationally efficient in generating and providing a compact yet comprehensive representation of normal layer behavior in fault-free AxDNNs. For each layer $l$, a statistical signature $S_l$ is computed as:

$$S_l = \{\mu_l, \sigma_l, Q_l, \rho_l\} \quad (1)$$

In Equation (1), $\mu_l$ and $\sigma_l$ capture the mean and standard deviation of each layer's weights, representing central tendency and variability. Quartile metrics $Q_l$ enhance robustness by mitigating the influence of outliers, a critical consideration in approximate settings. The sparsity pattern $\rho_l$, which quantifies the distribution of zero and non-zero weights, is particularly relevant in approximate computing, where weights near zero are frequently quantized to zero. Deviations in $\rho_l$ serve as sensitive indicators of structural inconsistencies or faults. These metrics collectively form a baseline for detecting runtime deviations. By confining the signature generation to the pre-deployment phase, EPSILON achieves a lightweight design, ensuring the framework is well-suited for resource-constrained edge hardware deployments.

*2) Layer-Importance Aware Detection:* EPSILON introduces a layer-importance-aware detection mechanism to account for the varying sensitivities of AxDNN layers to hardware faults. Different layers contribute unequally to the overall network performance, and faults in critical layers can disproportionately degrade model accuracy [34]. Early convolutional layers, which extract foundational features, are often more fault-sensitive than later layers that refine or combine high-level abstractions [35]. To capture this variability, we define an importance factor $\alpha_l$ for each layer $l$ as:

$$\alpha_l = \beta_p \cdot \gamma_s \quad \text{where } \beta_p \in [0,1], \gamma_s \in [0,1] \quad (2)$$

In Equation (2), $\beta_p$ denotes the positional importance of the layer, progressively decreasing from input to output layers. This reflects the observation that faults in earlier layers tend to propagate through the network, amplifying their impact. $\gamma_s$, on the other hand, captures the structural importance, incorporating attributes such as layer type, connectivity, and sparsity. For example, convolutional layers with high connectivity and feature extraction responsibility are assigned higher structural importance than activation layers or layers with significant sparsity [34]. Using $\alpha_l$, the fault detection thresholds are dynamically adjusted for each layer $l$ as:

$$T_l = (m + \alpha_l)\sigma_l \quad (3)$$

The threshold $T_l$ scales with both the intrinsic variability (standard deviation) of the layer's weights $\sigma_l$ and the importance factor $\alpha_l$ with the offset $m$. The ranges of $[0,1]$ for $\beta_p$ and $\gamma_s$ are normalized to avoid over-weighting any single component and ensure balanced contributions.

## C. Inference Phase

The EPSILON framework operates in two distinct stages during inference to ensure efficient and accurate fault detection and mitigation. This dual-stage approach leverages multi-exit architectures for early detection and confidence-based mitigation, followed by a more comprehensive statistical analysis when necessary. The overall approach is presented in Algorithm 1, and the following subsections explain these stages. The algorithm takes as input a multi-exit AxDNN model $F$, a confidence threshold $\gamma$, statistical signatures $\{S_l = (\mu_l, \sigma_l, Q_l, \rho_l)\}_{l=1}^L$, layer importance factors $\{\alpha_l\}_{l=1}^L$, and a data sample $x$, and outputs the predicted label along with a fault detection boolean.

*1) Primary Detection through Exit Confidence:* The first stage of the runtime phase focuses on leveraging the multi-exit architecture for rapid, confidence-based fault detection. For a given input $x$, each exit $i$ computes its prediction $F_i(x)$, and the confidence $conf_i$ is determined (see Algorithm 1, Lines 3–4) as follows:

$$conf_i = \max(F_i(x)) \quad \text{for exit } i \in \{1, \ldots, N\}. \quad (4)$$

Next, the framework checks whether any exit produces a confidence score exceeding the predefined threshold $\gamma$ (Algorithm 1, Line 5):

$$\exists i : conf_i > \gamma. \quad (5)$$

**Algorithm 1** Proposed EPSILON Framework

**Inputs**: *Multi-exit AxDNN: $F$*
  *Confidence threshold: $\gamma$*
  *Statistical signatures: $\{S_l = (\mu_l, \sigma_l, Q_l, \rho_l)\}_{l=1}^L$*
  *Layer importance factors: $\{\alpha_l\}_{l=1}^L$*
  *Input: $x$*
**Outputs:** *Prediction label, Fault detection boolean*

1 $list \leftarrow \emptyset$ // Store predictions
2 $all\_exits\_low\_confidence \leftarrow$ True
3 **for** $i = 1$ **to** $N$ **do**
   // N is the number of exits
4   $pred_i \leftarrow F_i(x)$  $conf_i \leftarrow \max(pred_i)$
5   **if** $conf_i > \gamma$ **then**
6     $all\_exits\_low\_confidence \leftarrow$ False **return** $(pred_i)$, *False*
7   **end**
8   $list.\text{Add}((pred_i))$
9 **end**
10 **if** $all\_exits\_low\_confidence$ **then**
   // Activate EPSILON framework
11   $fault\_detected \leftarrow$ False
12   **for** $l = 1$ **to** $L$ **do**
13     $W_l \leftarrow \text{GetLayerWeights}(F, l)$
     $T_l \leftarrow (m + \alpha_l)\sigma_l$
     // Pattern analysis
14     $pattern\_score \leftarrow \text{ComputePatternDeviation}(W_l, \rho_l)$
15     **if** $pattern\_score > T_l$ **then**
16       $fault\_detected \leftarrow$ True
17       **for** $w \in W_l$ **do**
18         **if** $|w - \mu_l| > T_l$ **then**
19           $w \leftarrow \text{FindNearestValid}(w, Q_l)$
20         **end**
21       **end**
22       $F \leftarrow \text{UpdateLayer}(F, l, W_l)$
23     **end**
24   **end**
   // Recompute prediction with corrected weights
25   $final\_pred \leftarrow F_N(x)$
26   **return** $(final\_pred)$, *fault_detected*
27 **end**

If this condition is met, the network outputs the corresponding prediction along with $False$ fault boolean value and bypasses further analysis (Algorithm 1, Line 6). This early exit mechanism minimizes computational overhead during regular operation, making it effective in scenarios where the predictions remain robust under faults.

*2) EPSILON Activation and Statistical Analysis:* When none of the exits achieve the required confidence ($conf_i \leq \gamma$ for all $i$), the EPSILON framework activates its fault detection mechanism (Algorithm 1, Line 10). This stage involves evaluating statistical deviations in layer-wise patterns to identify potential faults. Each layer $l$ is associated with a reference statistical signature $\rho_l$, and the current observed signature $\rho_{current}$ is compared using the pattern score (Algorithm 1, Line 14):

$$pattern\_score_l = \sum_{i=1}^{L} |\rho_l(i) - \rho_{current}(i)|. \quad (6)$$

The framework computes a dynamic threshold $T_l$ for each layer based on its reference standard deviation $\sigma_l$ and a tunable importance factor $\alpha_l$ (see Algorithm 1, Line 14). If $pattern\_score_l > T_l$, the layer is flagged as faulty (Algorithm 1, Line 16).

*3) Adaptive Mitigation Strategy:* EPSILON employs a targeted correction mechanism once a fault is detected in a layer. Each weight $w$ in the faulty layer is evaluated and corrected if it deviates significantly from its expected range (Lines 17–19 in Algorithm 1). The correction process adjusts weights to the nearest valid value in the set $Q_l$, representing acceptable weight ranges for the layer as:

$$w_{corrected} = \arg\min_{v \in Q_l} |w - v|. \quad (7)$$

where $v$ is candidate valid weight values. This step ensures that the layer's statistical integrity is restored while mitigating the impact of the detected faults. The corrected $w_{corrected}$ are updated to the network (Line 22 in Algorithm 1) and inference is performed with the updated weights (Line 25 in Algorithm 1) and the final prediction is returned along with $True$ fault boolean value (Line 26). The complete dual-stage approach enables an optimal balance between detection accuracy and computational efficiency.

### D. Theoretical Analysis and Implementation Insights

The EPSILON framework is designed for efficient implementation with minimal overhead. For a network with $N$ exits and $L$ layers, the framework requires $O(L)$ space for signature storage and $O(1)$ additional space for confidence tracking. In practice, the storage overhead of a typical network with $L = 50$ layers is less than 1KB per layer, making it feasible for resource-constrained environments. The detection and mitigation mechanisms operate in place, avoiding additional memory requirements during inference. The multi-exit architecture enables early termination in typical cases, reducing average-case computational complexity.

The effectiveness of EPSILON can be analyzed through the lens of statistical error bounds. Given a fault rate $p$ and layer importance $\alpha_l$, the probability of missed detection $P_{md}$ is bounded by:

$$P_{md} \leq \exp(-\alpha_l^2/2p) \qquad (8)$$

When combined with the multi-exit confidence check, the overall error probability is bounded by:

$$P_{error} \leq (1-\gamma)^N + \exp(-\alpha_l^2/2p), \qquad (9)$$

where $\gamma$ is the confidence threshold and $N$ is the number of exits. This bound ensures that critical layers maintain higher detection accuracy while allowing more flexibility in less crucial layers, aligning with the approximate computing paradigm.

## IV. EXPERIMENTAL RESULTS AND DISCUSSION

We comprehensively evaluate our proposed fault detection and mitigation framework across various network architectures, datasets, and approximate computing scenarios. The details are presented in the following subsections.

### A. Experimental Setup

**Dataset, Models and Baseline**: We employ four widely-used datasets: MNIST, CIFAR-10, CIFAR-100, and ImageNet-1k. By utilizing three popular DNN architectures for each dataset: VGG16, WideResNet, and MobileNet, providing a broad spectrum of network complexities and design philosophies. To compare the performance of EPSILON with state-of-the-art, we choose the approximate versions of Vanilla CNN (single-exit architecture) and MENDNet (multi-exit architecture) [16], denoted as AxVanDNN and AxMENDNet, respectively. It is worth mentioning that AxMENDNet also meets the JIT criteria while achieving state-of-the-art performance in AccDNNs [16], [8]. Furthermore, like EPSILON, MENDNet is built on top of the MENN, ensuring a fair comparison. We evaluate the performance of all selected algorithms using three key metrics: classification accuracy on held-out test sets to assess fault tolerance, inference latency for efficiency, and energy consumption for overall performance analysis. It is worth mentioning that AxDNN simulations are highly time-intensive, often requiring weeks or even months to complete, depending on the dataset size and the number of approximate multipliers used [36], [37]. This is a recognized limitation of EvoApprox8b and other hardware approximation libraries, which makes it challenging to train and simulate large-scale AxDNN models, particularly when working with extensive datasets and numerous approximate components. For instance, it took us approximately ***1,260 CPU hours*** to perform the experimentation for the datasets and DNN architectures we used to report the results in this paper.

**Approximate Computing**: To leverage the approximate computing in VGG16, WideResNet, and MobileNet models, we leverage a comprehensive selection of 8-bit signed approximate multipliers from Evoapproxlib [38]. We focus on multipliers as they dominate energy consumption in DNNs compared to other arithmetic units, such as adders [2]. The chosen eight different multipliers are KV6, KV8, KV9, KVP, L2J, L2H, L2N, and L2H. These multipliers represent various points along the energy-accuracy trade-off spectrum, ranging from mostly accurate but less energy-efficient designs to aggressive approximations offering substantial energy savings [39].

**Fault Model**: Hardware faults in approximate computing systems can manifest in various forms. In this work, we focus on permanent stuck-at faults, which are particularly relevant for approximate multipliers in neural network accelerators [35]. We consider both stuck-at-0 (SA0) and stuck-at-1 (SA1) faults that can occur in the weight storage and computation units of AxDNN accelerators. These faults typically manifest as bit flips in the binary representation of weights, affecting the least significant bits (LSBs) more frequently due to the nature of approximate computing [40]. The fault rate (FR) represents the percentage of weights affected in a given layer, where:

$$FR = \frac{N_{faulty}}{N_{total}} \times 100\%, \qquad (10)$$

where $N_{faulty}$ is the number of faulty weights and $N_{total}$ is the total number of weights in the layer. The spatial distribution of faults follows a random pattern within each layer, which aligns with real-world observations in approximate hardware [41]. This randomness is particularly important as it reflects realistic fault patterns in approximate multipliers and provides worst-case scenarios for robustness analysis.

To thoroughly evaluate fault tolerance, we implement a strategic fault injection methodology. Faults are introduced at four carefully selected fault points (FP): FP1, FP2, FP3, and FP4 within each network, representing critical stages in the computational pipeline. These FPs were specifically chosen based on their demonstrated impact on network performance, as established through state-of-the-art [16]. The FP4 is positioned at the first conv layer, FP1 at the final classifier exit, and FP2-FP3 equally spaced between remaining internal classifiers. At each location, we inject faults at three different FR (10%, 30%, and 50%), creating a comprehensive test matrix. These FPs and FRs aligned well with our theoretical analysis, which predicts a bounded error probability of misdetection as shown in Equation (8). For instance, with confidence threshold $\gamma = 0.7$ and $N = 4$ exits (i.e., FP1-FP4), the theoretical bound predicts maintainable accuracy even at 50% fault rate when layer importance $\alpha_l$ is sufficiently high (>0.7).

**Computing Environment**: Our implementation utilizes the AdaPT framework, built on Evoapproxlib for PyTorch, and employs Python 3.12, pyJoules, PyTorch 2.1, and CUDA 11.8, running on an NVIDIA RTX 2080 Ti GPU and an Intel Core i9 CPU with 64 GB of system RAM.

### B. EPSILON Performance Analysis: Fault Tolerance

Due to space constraints, this section reports experimental results focusing on SA1 faults for the CIFAR-10 dataset using three representative multipliers (KVP, L2J, and L2H), chosen to illustrate the spectrum of approximation impacts as shown in Table I. The threshold values $\gamma$ and $\tau$ are determined by analyzing the post-fault accuracy of AxMENDNet models, selecting the thresholds that yield the best average accuracy for each model-dataset pair. For EPSILON, only the accuracy threshold $\gamma$ is considered. These values are different for each dataset, architecture, and approximate multiplier and are provided on the project website[1]. Comprehensive results across the rest of the multipliers and datasets for both SA1 and SA0 can be found there as well. Table I compares EPSILON with AxVanDNN and AxMEND-Net under SA1 faults for the CIFAR-10 dataset. The average accuracy (last column) is computed over all fault points (FP1–FP4) and FRs to enable a fair comparison with state-of-the-art for MENNs in AccDNNs [16], which also reports average accuracy as a benchmark metric. The results demonstrate a consistently superior performance pattern for EPSILON, and on average, it outperforms AxVanDNN and AxMENDNet across all FP and FR for all model-dataset-multiplier combinations. For instance, in the VGG16 architecture the EPSILON maintains accuracy of **80.05%** averaged across all FPs and FRs vs. 61.23% and 76.19% achieved by AxVanDNN and AxMENDNet, respectively. Similarly, on MobileNet with an aggressive approximate multiplier L2H, EPSILON achieves almost 8% more average accuracy than AxMENDNet while maintaining a lead on all FPs at all FRs. *Such a performance difference in average accuracies demonstrates EPSILON's effectiveness in mitigating AxDNN faults and shows that adopting state-of-the-art AccDNN fault mitigation approaches (e.g., AxMENDNET and AxVanDNN)* ***does not*** *perform well for mitigating AxDNN faults.* WideResNet exhibits particularly strong resilience at FP2 and FP3, suggesting that its architecture provides natural fault tolerance at these locations. MobileNet benefits significantly from our EPSILON framework's mitigation strategies while showing more uniform degradation across FPs and maintaining higher accuracy than AxVanDNN and AxMENDNet.

TABLE I: Comparison of classification accuracy (%) of AxVanDNN, AxMENDNet, and EPSILON across different fault points (FP1–FP4) and fault rates (FR: 10%, 30%, 50%) using KVP, L2J, and L2H approximate multipliers (AppxMults) for VGG16, WideResNet (WideRN) and MobileNet architectures on the CIFAR-10 dataset under SA1 faults

| Models | AppxMults | Techniques | FP#1 | | | FP#2 | | | FP#3 | | | FP#4 | | | Avg. acc. % |
|---|---|---|---|---|---|---|---|---|---|---|---|---|---|---|---|
| | | | FR:10% | FR:30% | FR:50% | FR:10% | FR:30% | FR:50% | FR:10% | FR:30% | FR:50% | FR:10% | FR:30% | FR:50% | |
| VGG16 | KVP | AxVanDNN | 88.12 | 85.05 | 18.84 | 88.15 | 85.08 | 85.11 | 85.13 | 82.82 | 10.23 | 77.71 | 35.34 | 10.18 | 61.23 |
| | | AxMENDNet | 85.15 | 84.12 | 83.08 | 85.11 | 84.14 | 83.13 | 83.09 | 82.02 | 81.94 | 69.68 | 32.92 | 19.87 | 76.19 |
| | | EPSILON | 93.83 | 92.75 | 89.72 | 91.78 | 89.74 | 88.76 | 88.73 | 87.52 | 86.98 | 85.82 | 39.24 | 25.78 | **80.05** |
| | L2J | AxVanDNN | 86.52 | 82.45 | 16.84 | 86.55 | 82.48 | 82.51 | 82.53 | 79.82 | 9.23 | 74.21 | 32.84 | 9.18 | 58.93 |
| | | AxMENDNet | 79.55 | 78.52 | 77.48 | 79.51 | 78.54 | 77.53 | 77.49 | 76.42 | 75.34 | 63.18 | 26.42 | 13.37 | 65.19 |
| | | EPSILON | 85.23 | 84.15 | 83.12 | 85.18 | 84.14 | 83.16 | 83.13 | 81.92 | 80.81 | 78.32 | 30.74 | 14.28 | **71.54** |
| | L2H | AxVanDNN | 75.82 | 65.75 | 12.24 | 75.85 | 65.78 | 65.81 | 65.83 | 55.12 | 7.53 | 45.51 | 25.14 | 7.48 | 47.23 |
| | | AxMENDNet | 68.85 | 67.82 | 66.78 | 68.81 | 67.84 | 66.83 | 66.79 | 65.72 | 64.64 | 52.48 | 22.72 | 10.67 | 54.99 |
| | | EPSILON | 74.53 | 73.45 | 72.42 | 74.48 | 73.44 | 72.46 | 72.43 | 71.22 | 70.11 | 67.62 | 27.04 | 11.08 | **62.24** |
| WideRN | KVP | AxVanDNN | 85.82 | 70.48 | 34.24 | 88.34 | 85.28 | 82.25 | 85.31 | 82.73 | 80.14 | 73.32 | 62.05 | 32.72 | 71.47 |
| | | AxMENDNet | 78.85 | 77.73 | 76.68 | 78.62 | 77.45 | 76.75 | 76.58 | 75.34 | 74.82 | 67.68 | 26.42 | 13.23 | 65.85 |
| | | EPSILON | 88.32 | 87.28 | 86.25 | 88.18 | 87.52 | 86.48 | 86.43 | 85.87 | 82.92 | 63.24 | 46.32 | 11.12 | **74.41** |
| | L2J | AxVanDNN | 83.22 | 68.88 | 32.64 | 85.74 | 82.68 | 79.65 | 82.71 | 80.13 | 77.54 | 70.72 | 59.45 | 30.12 | 69.87 |
| | | AxMENDNet | 76.25 | 75.13 | 74.08 | 76.02 | 74.85 | 74.15 | 73.98 | 72.74 | 72.22 | 65.08 | 24.82 | 11.63 | 63.25 |
| | | EPSILON | 85.72 | 84.68 | 83.65 | 85.58 | 84.92 | 83.88 | 83.83 | 83.27 | 80.32 | 60.64 | 43.72 | 9.52 | **71.81** |
| | L2H | AxVanDNN | 72.52 | 58.18 | 22.94 | 75.04 | 72.98 | 69.95 | 72.01 | 65.43 | 57.84 | 45.02 | 35.75 | 20.42 | 55.17 |
| | | AxMENDNet | 65.55 | 64.43 | 63.38 | 65.32 | 64.15 | 63.45 | 63.28 | 62.04 | 61.52 | 49.38 | 22.12 | 8.93 | 52.55 |
| | | EPSILON | 75.02 | 73.98 | 72.95 | 74.88 | 74.22 | 73.18 | 73.13 | 72.57 | 69.62 | 49.94 | 38.02 | 7.82 | **61.11** |
| MobileNet | KVP | AxVanDNN | 85.03 | 82.23 | 13.18 | 85.17 | 82.21 | 82.15 | 80.98 | 68.02 | 26.23 | 77.02 | 52.82 | 23.32 | 62.78 |
| | | AxMENDNet | 80.92 | 79.87 | 78.85 | 80.78 | 79.72 | 78.68 | 78.84 | 77.42 | 76.75 | 63.73 | 20.14 | 8.82 | 66.29 |
| | | EPSILON | 85.62 | 84.58 | 83.55 | 85.64 | 84.57 | 83.59 | 82.92 | 80.48 | 77.73 | 77.54 | 35.68 | 27.42 | **73.36** |
| | L2J | AxVanDNN | 82.43 | 79.63 | 12.58 | 82.57 | 79.61 | 79.55 | 78.38 | 65.42 | 24.63 | 74.42 | 50.22 | 21.72 | 60.18 |
| | | AxMENDNet | 78.32 | 77.27 | 76.25 | 78.18 | 77.12 | 76.08 | 76.24 | 74.82 | 74.15 | 61.13 | 18.54 | 7.22 | 63.69 |
| | | EPSILON | 83.02 | 81.98 | 80.95 | 83.04 | 81.97 | 80.99 | 80.32 | 77.88 | 75.13 | 74.94 | 33.08 | 25.82 | **70.76** |
| | L2H | AxVanDNN | 71.73 | 68.93 | 10.88 | 71.87 | 68.91 | 68.85 | 67.68 | 54.72 | 20.93 | 63.72 | 39.52 | 18.02 | 51.48 |
| | | AxMENDNet | 67.62 | 66.57 | 65.55 | 67.48 | 66.42 | 65.38 | 65.54 | 64.12 | 63.45 | 50.43 | 15.84 | 5.52 | 52.99 |
| | | EPSILON | 72.32 | 71.28 | 70.25 | 72.34 | 71.27 | 70.29 | 69.62 | 67.18 | 64.43 | 64.24 | 28.38 | 21.12 | **60.06** |

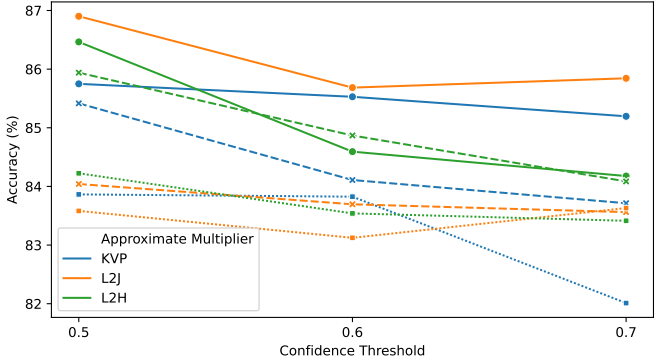

Fig. 1: Impact of confidence threshold ($\gamma$) on accuracy for approximate multipliers KVP, L2J, and L2H on WideRN with FP1 with FR of 10% on SA1 for CIFAR-10 dataset

Our obtained empirical results also validate the proposed theoretical error-bound mathematical model as shown in Equation (9). For example, EPSILON achieved 88.32% accuracy at FP1 and maintained >82% accuracy through FP3 for WideRN with KVP, even under challenging fault scenarios. This accuracy at FP1 aligns with our theoretical bound which predicts a minimum accuracy of 81.45% under these conditions $((1-0.7)^4 + \exp(-0.7^2/2(0.5)) \approx 0.1855)$. The observed performance degradation at FP4 (dropping to 11.12% at 50% fault rate) also follows our mathematical model, demonstrating the framework's predictable behavior across different fault scenarios while validating our theoretical foundations.

[1] https://github.com/dependable-cps/EPSILON

### C. EPSILON approximation levels vs. fault tolerance

The relationship between approximation levels and fault tolerance reveals an interesting trend. As we progress from KVP to L2H, representing increasing levels of approximation, we observe a predictable decrease in baseline accuracy. However, the EPSILON framework consistently demonstrates smaller accuracy drops than AxVanDNN and AxMENDNet. This performance gap becomes particularly pronounced at higher FRs, highlighting EPSILON's effectiveness in managing the combined challenges of approximation and fault tolerance. The impact of fault location emerges as a critical factor in performance degradation patterns. Our analysis reveals that while FP1 through FP3 demonstrate relatively stable performance under EPSILON, FP4 consistently presents more challenging conditions for all approaches. The observed accuracy degradation at **FP4** can be attributed to the critical role of early convolutional layers in feature extraction. Unlike FP1 (final classifier exit) or FP2-FP3 (internal exits), faults injected at FP4 disrupt the initial feature maps, leading to a cascading error propagation throughout the network. Since convolutional layers apply shared filters across the entire input, even minor faults in FP4 can introduce widespread distortions, significantly impacting subsequent layers. This amplification effect makes early-layer faults particularly detrimental compared to faults occurring in deeper layers, where redundancy and multi-exit mechanisms provide more opportunities for error compensation. Furthermore, while the EPSILON framework effectively mitigates faults in later layers using statistical signatures and confidence-based exits, its mitigation strategies are inherently less effective at FP4. Since faults at this stage corrupt the foundational features rather than high-level representations, the network struggles to recover from the introduced errors, resulting in a more pronounced accuracy drop. This aligns with our empirical findings, where accuracy at FP4 degrades more severely under high fault rates compared to FP1-FP3. This trend holds across different architectures and approximation levels, though the magnitude of impact varies. The EPSILON framework shows particular strength in maintaining accuracy at the critical fault points FP1-FP3, where preserving network functionality is crucial for overall performance.

### D. EPSILON Sensitivity to $\gamma$ and $\alpha_l$

The confidence threshold $\gamma$ and layer importance factor $\alpha_l$ in (Algorithm 1, Line 5) and Equation (3) are critical in defining the robustness and efficiency of the EPSILON framework. The $\gamma$ serves as a sensitivity threshold for early fault detection in AxDNN. Low $\gamma$ values enhance sensitivity to minor deviations but increase false positives, incurring higher computational overhead. In contrast, high $\gamma$ values focus on significant faults, potentially overlooking critical early-stage errors. Based on empirical evaluation, $\gamma = 0.5$ balances fault detection accuracy and false positive rates for CIFAR-10. The layer-specific scaling factor $\alpha_l$ adjusts fault mitigation based on the sensitivity and importance of each layer. Uniformly low $\alpha_l$ values simplify the fault handling process but fail to account for vulnerabilities in deeper layers prone to cumulative errors. Conversely, high $\alpha_l$ values prioritize critical layers, though at the cost of increased complexity and resource allocation challenges. An adaptive $\alpha_l$ configuration ensures optimal fault resilience across layers. The offset factor $m$ with value 3 ensures a baseline level of fault tolerance, allowing the mechanism to remain robust even for layers with low importance. Fig. 1 illustrates the impact of $\gamma$ and Fig. 2 illustrates the impact of $\alpha$ on WideRN with faults injected at FP1 with FR 30% on CIFAR-10, showing averaged accuracy across FP1 at FR of 10% for three approximate multipliers on the VGG16 model by varying the $\gamma$ and $\alpha_l$ at a time while keeping other(s) fixed. Fig. 1 demonstrate a notable trend wherein an increase in the confidence threshold results in a corresponding reduction in model accuracy. This phenomenon can be attributed to the fact that elevating the confidence threshold imposes stricter criteria for accepting predictions from earlier exit points within the MENN architecture. As a consequence, a larger proportion of inputs are required to propagate through deeper layers of the network. During this propagation, any hardware faults present in earlier layers are exacerbated due to the cumulative effect of subsequent multiplication and addition operations. Furthermore, if hardware faults occur within the later layers, they directly influence the final output, thereby contributing to a decline in overall accuracy. These findings underscore the importance of carefully selecting the confidence threshold value, as it represents a critical hyperparameter in MENNs. The optimal threshold is contingent upon factors such as the dataset, network architecture, and the specific characteristics of the approximate multipliers employed. Section IV.C further discusses the sensitivity of EPSILON to parameters $\gamma$ and $\alpha$. The impact on other multipliers and datasets can be found on the project website [1].

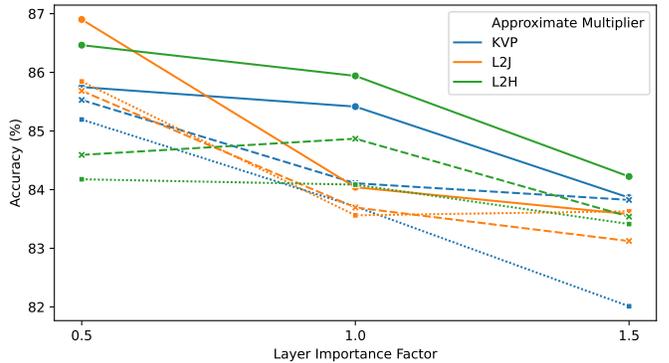

Fig. 2: Impact of layer importance factor ($\alpha_l$) on accuracy for approximate multipliers KVP, L2J, and L2H on WideRN with FP1 with FR of 10% on SA1 for CIFAR-10 dataset

### E. EPSILON Efficiency Analysis: Inference Latency

In this section, we evaluate the computational efficiency of our approach through inference time analysis across all datasets and representative approximate multipliers as shown in Fig 3. Our results demonstrate that in addition to superior fault tolerance, as shown in Section IV-B, EPSILON outperforms the state-of-the-art works in terms of inference times across all datasets. The results for the rest of the multipliers are available on the project website[1]. For instance, on ImageNet with KVP, EPSILON achieves an average inference time of 220ms, compared to AxMENDNet's 228.45ms and vanilla AxDNN's 245.62ms. This difference is because AxVanDNN traverses through the complete network every time, while EPSILON and AxMENDNet terminate early if they meet the exit criteria, i.e., $\gamma$ for EPSILON while $\gamma$ and $\tau$ for AxMENDNet. This observation

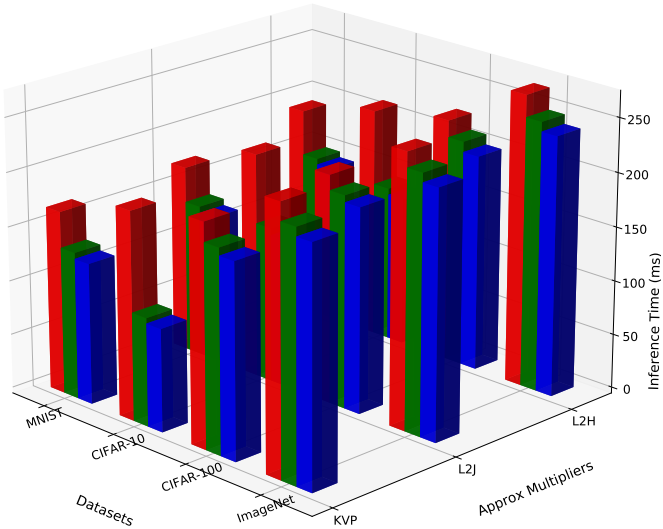

Fig. 3: Average inference time (ms) for AxVanDNN (red), AxMEND-Net (green), and EPSILON (blue) with the representative multipliers on MNIST, CIFAR-10, CIFAR-100 and ImageNet datasets for SA1

aligns with the existing studies [8], [16]. However, the computational efficiency of EPSILON over AxMENDNet becomes particularly apparent when examining the underlying mechanisms. While both frameworks utilize a multi-exit architecture, AxMENDNet's requirement to compute both confidence score ($\gamma$) and confidence threshold ($\tau$) for each inference introduces additional overhead both in terms of inference time and energy consumption. Our results indicate that this dual computation accounts for approximately 28% of AxMEND-Net's total processing time. In contrast, EPSILON's streamlined approach of using only the confidence score ($\gamma$) combined with pre-computed statistical signatures reduces computational overhead while maintaining robust fault detection capabilities. This architectural difference explains the observed performance gap in inference times: on average, EPSILON consistently achieves **23.5%** faster processing than AxMENDNet across MNIST, CIFAR-10, CIFAR-100, and ImageNet datasets with KVP, L2J, and L2H multipliers, as shown in Fig 3. This efficiency advantage is most pronounced on the CIFAR-10 dataset, where EPSILON requires only 95ms compared to AxVanDNN's 192.16ms with the KVP multiplier, while similar trends hold for other datasets. Interestingly, the performance gap widens with more aggressive approximation levels, as seen with L2H, where EPSILON maintains relatively stable inference times (ranging from 145ms on MNIST to 240ms on ImageNet) while AxVanDNN and AxMENDNet show more significant slowdowns (reaching up to 268.95ms and 248.72ms respectively on ImageNet). This trend suggests that our framework's multi-exit architecture and efficient statistical analysis effectively balance the overhead of fault detection with the benefits of early termination. The consistent inference time advantages across different approximate multipliers also indicate that EPSILON's statistical pattern-based approach scales well with varying levels of hardware approximation, making it particularly suitable for real-world deployments where both reliability and computational efficiency are critical considerations.

*F. EPSILON Performance Analysis: Energy Efficiency*

This section presents the energy consumption results (measured using the pyJoules library) of the EPSILON, AxMENDNet, and AxVanDNN frameworks, as shown in Fig. 4, across all datasets with SA1 faults on the KVP approximate multiplier. Similar trends are observed for the remaining multipliers, corresponding to their approximation levels [1]. Each inference is repeated 100 times to ensure statistical significance. For better representation, we normalized the results against the peak energy consumption of AxMENDNet on ImageNet-1k as the baseline. As shown in Fig. 4, EPSILON demonstrates superior energy efficiency, operating at **52.9%–80.7%** of the baseline across all datasets. This represents a significant improvement over both AxMENDNet (77.7%–100%) and AxVanDNN (79.7%–97.2%). This relative analysis reveals that EPSILON consistently outperforms AxMENDNet by an average margin of **29.4%** and AxVanDNN by **27.8%**. This substantial efficiency gain can be attributed to EPSILON's adaptive early-exit strategy and targeted fault detection mechanism, which minimizes unnecessary computations while maintaining reliability. These results position EPSILON as a promising solution for energy-efficient approximate computing systems, particularly in resource-constrained environments.

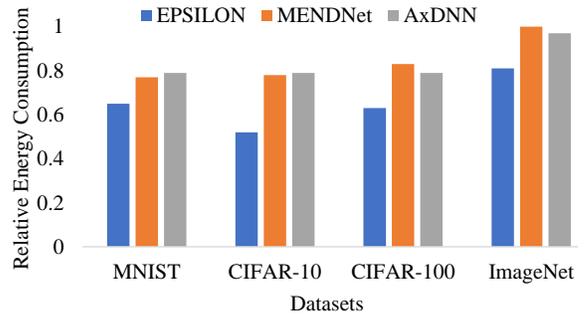

Fig. 4: Comparison of relative energy consumption across algorithms and datasets on SA1 faults for KVP approximate multiplier

## V. CONCLUSION

This paper presented EPSILON, a novel framework for enhancing the reliability of Approximate Deep Neural Networks (AxDNNs) in energy-constrained environments. By leveraging multi-exit neural network architecture and statistical pattern analysis, EPSILON effectively distinguishes between intentional approximation effects and permanent hardware faults while maintaining computational efficiency. Our comprehensive evaluation across multiple datasets and network architectures demonstrates EPSILON's superior fault tolerance, inference time, and energy efficiency performance, outperforming other state-of-the-art approaches. In the future, we plan to evaluate the effectiveness of EPSILON against larger models for different model architectures trained on large datasets.

## VI. ACKNOWLEDGMENTS

This material is based upon work supported by the National Science Foundation (NSF) under Award Numbers: CCF-2323819. Any opinions, findings, conclusions, or recommendations expressed in this publication are those of the authors and do not necessarily reflect the views of the NSF.